\documentclass[twocolumn,showpacs,preprintnumbers]{revtex4}
\usepackage{amssymb}
\usepackage{amsfonts}
\usepackage{amsmath}
\usepackage{graphicx}
\usepackage{dcolumn}
\usepackage{bm}

\begin{document}

\title{Fractional quantum Hall effect without energy gap}
\author{S. S. Murzin, S. I. Dorozhkin, G. E. Tsydynzhapov and V. N. Zverev}
\affiliation{Institute of Solid State Physics RAS, 142432, Chernogolovka, Moscow
District, Russia }

\begin{abstract}
In the fractional quantum Hall effect regime we measure diagonal ($\rho _{xx}
$) and Hall ($\rho _{xy}$) magnetoresistivity tensor components of
two-dimensional electron system (2DES) in gated GaAs/Al$_{x}$Ga$_{1-x}$As
heterojunctions, together with capacitance between 2DES and the gate. We
observe 1/3- and 2/3-fractional quantum Hall effect at rather low magnetic
fields where corresponding fractional minima in the thermodynamical density
of states have already disappeared manifesting complete suppression of the
quasiparticle energy gaps.
\end{abstract}

\maketitle

Energy gaps in the quasiparticle excitation spectrum play important role for
explanation of both integer and fractional quantum Hall effects (QHE) in the
most of theoretical constructions (for review see Ref.~\cite{Prange}).
However, for the integer QHE only the mobility gap is required~\cite{Laugh}
which in general may appear in the absence of the energy gap. Existence of
the integer QHE without Landau quantization and corresponding modulation of
electron density of states has been predicted~\cite{Khm} on the base of the
scaling approach for the case of low magnetic fields ($\omega _{c}\tau \ll 1$%
, where $\omega _{c}$ is the cyclotron frequency and $\tau $ is the electron
relaxation time) when the cyclotron energy gaps are completely suppressed by
disorder. This effect has not been found so far, probably, because of
extremely low temperatures required for its manifestation~\cite{Huck}.
However, the observation of the integer QHE in thick heavily Si-doped n-type
GaAs layers with three-dimensional energy spectrum in the extreme quantum
limit of applied magnetic field~\cite{Mur}, strongly supports the idea that
the integer QHE is not necessarily related to the energy gap. Since the
scaling treatment of both integer~\cite{Khm,Prb} and fractional~\cite%
{Lut,Dolan,BD,Pr99} QHE relies on the values of magnetoconductivity tensor
components irrespectively to features of quasiparticle energy spectrum a
question arises: is the fractional QHE possible in the absence of the energy
gap by analogy with the integer QHE? More specifically, the question
addressed in this paper is whether a disorder, which is known to suppress
the fractional QHE and the energy gap, suppresses them simultaneously or the
energy gap completely vanishes before the fractional QHE (and the mobility
gap of quasiparticles) disappears.

\begin{figure}[tb]
\includegraphics[width=8.5cm,clip]{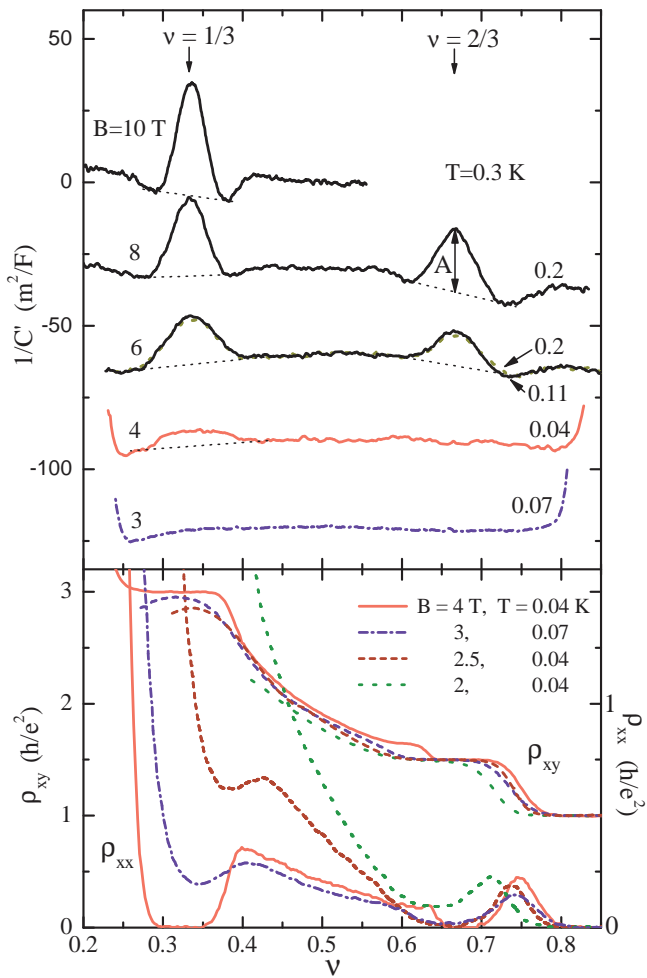}
\caption{(color online). (upper panel) Variation of the inverse capacitance $%
1/C^{\prime }=1/C-(5568-90\protect\nu )$, and (lower panel) the diagonal ($%
\protect\rho _{xx}$) and Hall ($\protect\rho _{xy}$) resistivities (both
given in units of $h/e^{2}$) versus filling factor $\protect\nu $ tuned by
the gate voltage $V_{g}$ at different magnetic fields and temperatures,
shown in the figures, for sample 1. The capacitance curves measured at
different magnetic fields are vertically shifted, the left scale corresponds
to 10 T curve. Dotted straight lines in upper panel were used as baselines
for determining peak heights $A$.}
\label{C1}
\end{figure}

Here we present experimental data that indicate that the fractional QHE can
survive in a rather weak magnetic field where the fractional features in the
energy spectrum have already vanished. In the fractional QHE regime, we
measure diagonal ($\rho _{xx}$) and Hall ($\rho _{xy}$) magnetoresistivity
tensor components of two-dimensional electron system (2DES) in gated GaAs/Al$%
_{x}$Ga$_{1-x}$As heterojunctions, together with capacitance between 2DES
and the gate. Inverse value of the capacitance per unit area $C$ depends
linearly on $d\mu /dn$~\cite{Sm}, the inverse value of the thermodynamical
density of states (here $n$ is the areal electron density and $\mu $ is the
chemical potential of 2DES). The energy gap leads to a jump of the chemical
potential $\mu $ at a fractional filling factor~\cite{Halp} and, therefore,
to a peak in $d\mu /dn$. This peak has been found in experiments on electric
field penetration through 2DES~\cite{Eisen} and capacitance measurements~%
\cite{Dor}. Vanishing of the peak implies disappearance of the corresponding
minimum in the thermodynamical density of states, i.e., collapse of the
energy gap.

\begin{figure}[t]
\includegraphics[width=8.5cm,clip]{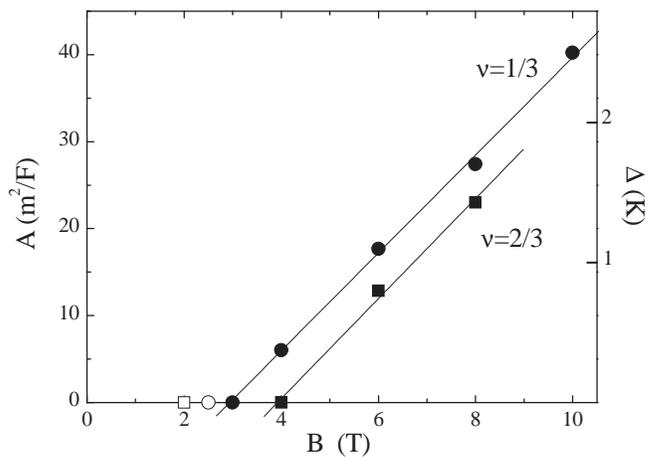}
\caption{Amplitude of the peaks $A$ in the inverse capacitance $1/C$ as a
function of magnetic field (solid symbols). Open symbols show minimal field
values where development of the fractional QHE is observed. Estimation of
the gap values are shown by the right scale.}
\label{A}
\end{figure}

We study two gated samples produced from two different wafers with single
remotely doped GaAs/Al$_{x}$Ga$_{1-x}$As heterojunction. They have an
undoped Al$_{x}$Ga$_{1-x}$As spacer of 70 and 40~nm width and the distance
between the gate and the heterojunction is 580 and 110 nm for sample 1 and
2, respectively. At zero gate voltage, the samples had electron densities $%
n_{1}=1.4\times 10^{11}$ and $n_{2}=1.8\times 10^{11}$~cm$^{-2}$ and
mobilities $\mu _{1}=1.2\times 10^{6}$ and $\mu _{2}=6.8\times 10^{5}$~cm$%
^{2}$/Vs. The density varies linearly with the gate voltage practically down
to the point of complete depletion of 2DES. The Hall bar geometry is used to
measure the Hall and diagonal resistivities. To measure capacitance the gate
voltage $V_{g}$ is modulated by low frequency (7-9 Hz) ac voltage and the
two components of ac current between the gate and 2DES are measured. The
component shifted by $90^{\circ }$ relative to the modulation voltage is
proportional to the capacitance. To obtain correct amplitude and width of
the fractional peak, we use the modulation amplitude below 15 and 4~mV for
sample 1 and 2, respectively. The in-phase component appears only at very
low values of $\sigma _{xx}$ due to the resistive effects in 2DES~\cite{Zh}.
These effects prevent charging of 2DES and lead to a drastic decrease of the
out-of-phase\ current component. To measure capacitance in the fractional
QHE regime accurately it is necessary to have values of $\sigma _{xx}$ that
are not too small, which sets a lower limit on the temperature of
measurement depending on the magnetic field value. Just above this limit,
temperature dependence of the capacitance is weak, so the peak amplitude
measured at such conditions is expected to be close to the amplitude at zero
temperature. Note, that in the range of magnetic field that is of special
interest for us, where the energy gap and QHE successively disappear, the
measurements are limited from below by the lowest temperature we can achieve.

\begin{figure}[tb]
\includegraphics[width=8.5cm,clip]{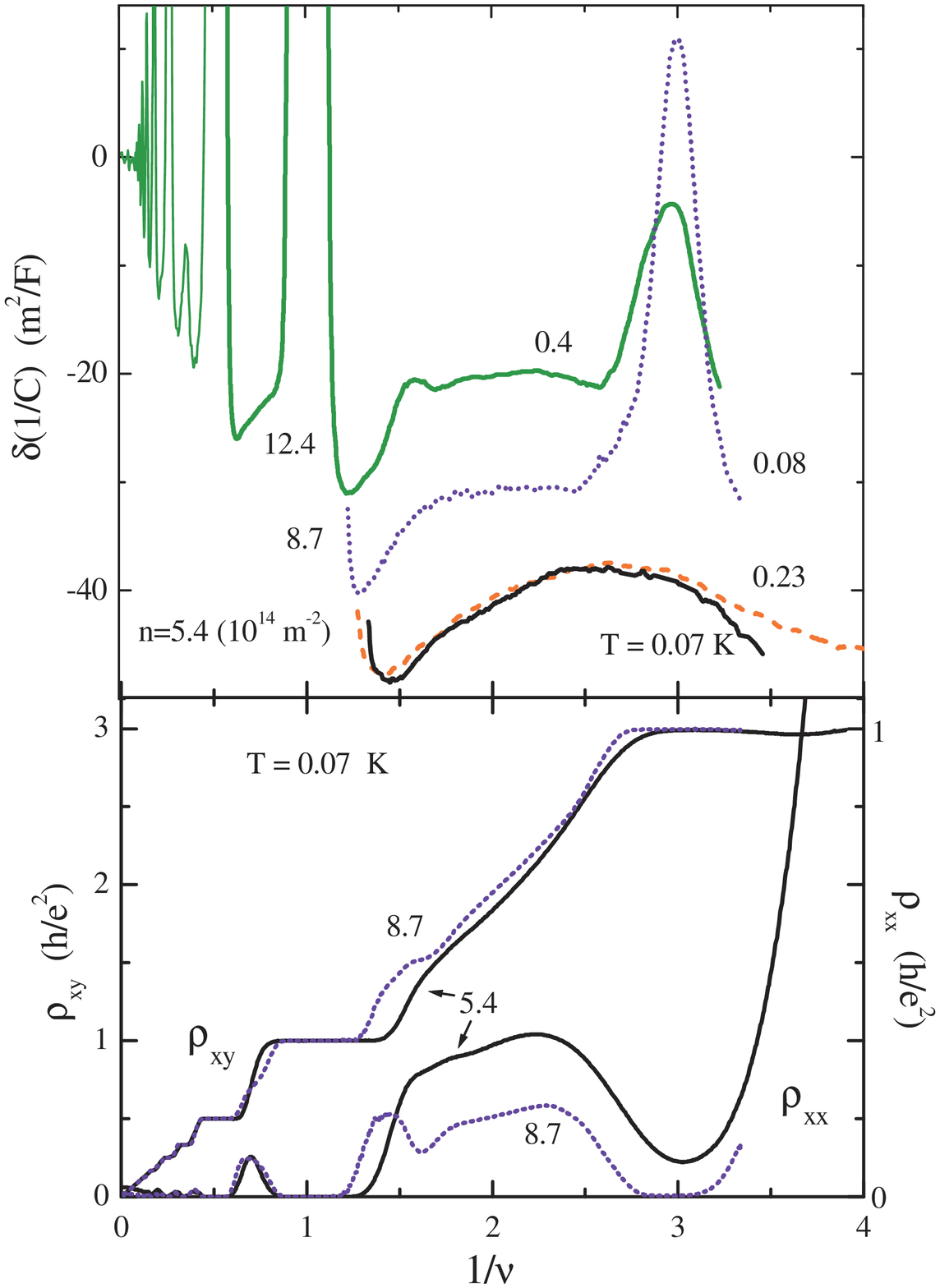}
\caption{(color online). (upper panel) Variation of the inverse capacitance $%
\protect\delta (1/C)=1/C-1/C_{B=0}$ and (lower panel) magnetotransport data
for sample 2 versus the inverse filling factor varied by magnetic field for
different electron densities $n$ (in units $10^{14}$ m$^{-2}$) and
temperatures (in K) shown in the figures. $\protect\delta (1/C)$-curves for
different electron densities are shifted vertically for clarity. The left
scale corresponds to $n=12.4\times 10^{14}{m}^{-2}$ curve. The total inverse
capacitance value $1/C\approx 1300$~m$^{2}$/F.}
\label{C2}
\end{figure}

\begin{figure}[t]
\includegraphics[width=8.5cm,clip]{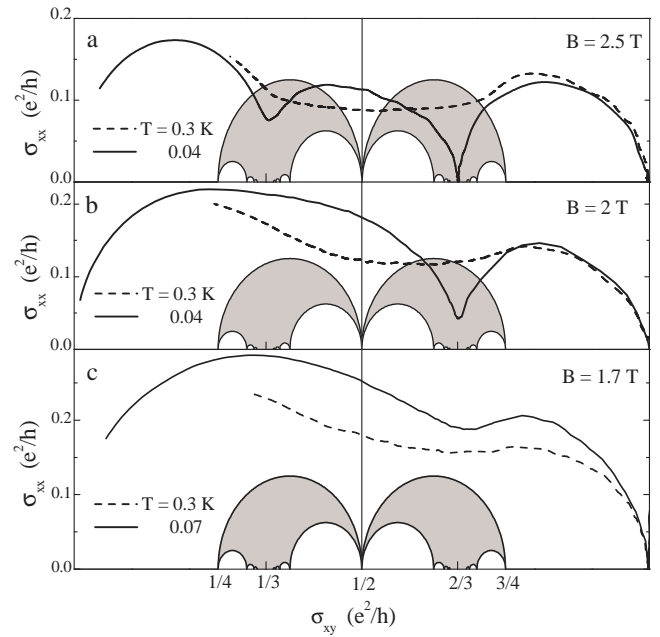}
\caption{Magnetotransport data representation in the $\protect\sigma _{xy}-%
\protect\sigma _{xx}$ plane. Theoretical regions of conductivities
converging to the 1/3- and 2/3- fractional QHE states when temperature goes
to zero are shown by left and right grey areas, respectively. The dotted and
solid lines present experimental conductivity traces obtained by sweeping
gate voltage for $T=0.3$~K and the base temperature, respectively. The
traces in panels (a), (b), and (c) correspond to indicated magnetic fields.}
\label{D}
\end{figure}

Experimental results for sample 1 are shown in Fig.~\ref{C1} as a function
of filling factor $\nu =nh/eB$, tuned by the gate voltage. The upper panel
shows how the peaks in the inverse capacitance related to energy gaps in the
1/3- and 2/3-fractional QHE states damp with decreasing magnetic fields. In
order to stress peaks in $1/C(\nu )$ linear function of $\nu $ with the
slope of $-90$ m$^{2}/$F is subtracted from all experimental data and
resulting curves are shifted vertically for clarity. The data are measured
under conditions at which resistive effects do not affect the results yet,
but temperature dependence of the capacitance is already weak. It is
illustrated by comparison of curves measured in 6 T field at 0.11~K and
0.2~K. This supports our expectations that the magnetocapacitance data
approximately reflect their zero-temperature behavior. In the lower panel,
the diagonal $\rho _{xx}$ and Hall resistivity $\rho _{xy}$ are shown.
Comparison of the magnetocapacitance and magnetotransport data for $B=4$~T
and $T=0.04$~K clearly shows that a well-developed QHE exists at $\nu =2/3$
while the corresponding peak on the magnetocapacitance curve at $\nu =2/3$
is absent. Moreover, the QHE features in the magnetotransport survive down
to substantially lower magnetic fields (2 T). This is the key observation
which indicates the existence of QHE at $\nu =2/3$ in the absence of the
energy gap. Similarly, the 1/3-QHE peak on the capacitance curves disappears
at $B=3$\ T~\cite{comment} while QHE\ features are observed at $B=3$\ T and
even at 2.5 T in the magnetotransport data. The amplitudes of the peaks
(determined as shown in Fig.~\ref{C1}) as a function of the magnetic field
(see Fig.~\ref{A}) follow straight lines that go to zero at threshold
magnetic fields, and QHE is still observed below the thresholds. The energy
gaps $\Delta $ can be estimated~\cite{comment2} from the peak areas $S$ in
Fig.~\ref{C1}. Corresponding values are given by the right axis in Fig.~\ref%
{A}.

Fig.\ref{C2} shows $\delta (1/C)=1/C(B,V_{g})-1/C(B=0,V_{g})$ (upper panel), 
$\rho _{xx}$ and $\rho _{xy}$ (lower panel) as a function of the inverse
filling factor $1/\nu =eB/nh$ varied by sweeping magnetic field at different
electron densities for sample 2. At the largest electron density ($%
n=12.4\times 10^{14}$~m$^{-2}$) $\delta (1/C)$ shows a well-pronounced peak
at $\nu =1/3$ and weak peak at $\nu =2/3$ which are not affected by the
resistive effects in contrary to the huge peaks at integer filling factors.
Strong rise of the $\nu =1/3$ peak at $n=8.7\times 10^{14}$~m$^{-2}$and $%
T=0.08$~K occurs due to the resistive effects and is accompanied by
appearance of a peak in the in-phase current component which constitutes
about 10\% of the total capacitive current. The 1/3 peak in $\delta (1/C)$
completely disappears at $n=5.4\times 10^{14}$m$^{-2}$ while well developed
Hall plateau and deep minimum in the magnetoresistance are still observed.
Similar behavior is observed for $\nu =2/3$ where the peak disappears at $%
n=8.7\times 10^{14}$~m$^{-2}$\ but 2/3-QHE features survive.

Disappearance of the energy gap at a threshold magnetic field is established
in different experiments~\cite{Tsui,Dor} and is attributed to a disorder
inevitably present in real samples. The role of disorder in the fractional
QHE is a very complicated problem which, at the moment, is only partly
addressed on the microscopic level (for the case of the short-range
fluctuations of potential see, for example, results of recent numerical
calculations~\cite{Wan} and references therein). In the selectively doped
GaAs/AlGaAs single heterojunctions, dominating potential fluctuations are
generally accepted to have a long-range character because charged impurities
are located behind the spacer, which thickness $d$ is larger than the
inverse Fermi wave vector. For filling factors smaller than unity this is
equivalent to condition that magnetic length $l_{B}\ll d$. Potential
fluctuations lead to electron density fluctuations. Quantitative analysis of 
$d\mu /dn$ for such heterojunctions was proposed in Ref.~\cite{Efros} in
so-called narrow gap approximation. It assumes that characteristic energy $%
W\sim \sqrt{N_{i}}e^{2}/\varepsilon $ of smooth random potential is much
larger than the energy gap $\Delta $. Here $N_{i}$ is the average areal
density of charged donors in the doped layer which is close to the areal
density of 2D electrons at zero gate voltage and $\varepsilon $ is the
dielectric constant. When average electron density is close to the value $%
n_{f}$ corresponding to a fractional QHE, a sample is expected to be divided
into areas of compressible phase with $n<n_{f}$ and $n>n_{f}$ separated by
strips of incompressible phase with $n=n_{f}$. In this case, the peaks in
magnetocapacitance measurements are related to the incompressible strips,
nevertheless, values of the gap that are derived from them are expected to
be close to the gap of a pure system $\Delta _{p}\propto B^{1/2}$~\cite%
{Efros}. However, there is additional validity condition of this result: the
strip width $d_{str}$ cannot be smaller than the magnetic length $l_{B}$.
Since $d_{str}\propto \Delta _{p}^{1/2}\propto B^{1/4}$ and $l_{B}\propto
B^{-1/2}$\ condition $d_{str}>l_{B}$ fails at weak magnetic fields and the
incompressible strips should disappear leading to disappearance of the
capacitance peak. In this context our observation of different thresholds
for the capacitance peaks and for the fractional QHE imply that the presence
of incompressible strips is not necessary condition for the fractional
quantization. Note that in our case $W\sim 50$~K and condition $W\gg \Delta $
of the narrow gap approximation is fulfilled in the entire range of magnetic
fields in question.

Note that within the composite fermion approach~\cite{Jain}, our observation
resembles the prediction~\cite{Khm} for the integer QHE. Indeed, the absence
of the fractional features in the capacitance data leads to the conclusion
that the Landau quantization\ in the composite fermion energy spectrum is
completely smeared out by disorder.

According to the scaling theory, at infinite coherence length, integer and
fractional QHE states together with the insulating state produce complete
set of possible ground states for 2DES in magnetic field. Particular ground
state can be reached only if the magnetoconductivity tensor components at a
finite coherence length belong to a respective region in the $\sigma
_{xy}-\sigma _{xx}$ plane. The corresponding regions for 1/3 - and 2/3 -
fractional quantum Hall effect states~\cite{Dolan,BDD} are shown in Fig.~\ref%
{D} by grey areas. In these regions, points $(\sigma _{xy},~\sigma _{xx})$
flow to $(1/3,~0)$ or to $(2/3,~0)$ when temperature goes to zero. These
areas are upper-bounded by two semicircles connecting point $(1/2,~0)$ with
points $(1/4,~0)$ and $(3/4,~0)$. The semicircles are qualitatively similar
to the phase boundaries of Ref.~\cite{KLZ} in $\rho _{xy}-\rho _{xx}$ plane.
Our experimental data are in good quantitative agreement with the
theoretical prediction~\cite{Dolan,BDD}. If conductivity traces measured
with sweeping gate voltage enter the grey areas at $T=0.3$~K, the base
temperature data demonstrate either a well pronounced QHE state (2/3 in Fig.~%
\ref{D}a) or a tendency to converge to such state (1/3 in Fig.~\ref{D}a and
2/3 in Fig.~\ref{D}b). Otherwise lowering of the temperature results in
further deviation of the experimental curves from the grey areas (Fig.~\ref%
{D}c and left part of Fig.~\ref{D}b).

In summary, we have observed the 1/3 and 2/3 fractional QHE under conditions
when magnetocapacitance data do not show any sign of the energy gap.
Satisfactory quantitative agreement have been found with theoretical values
of high-temperature magnetoconductivity tensor components corresponding to
disappearance of these fractional QHE states.

The authors are grateful to MBE group from Max-Planck-Institut f\"{u}r Festk%
\"{o}rperforschung, Stuttgart, Germany for producing the samples. This work
was supported by the \emph{Russian Foundation for Basic Research }and \emph{%
\ INTAS }(S.I.D.)


\begin{thebibliography}{99}
\bibitem{comment} Capacitance data for $B=3$ T were obtained only above 0.07
K due to refrigerator performance issues.
\bibitem{comment2} $\Delta \approx S/3\times e^{3}B/h$. Here we take into
account that the quasiparticle energy gap $\Delta $ is equal to 1/3 of the
chemical potential jump which is due to the fact that for the fractional QHE
states under consideration addition or removal of one electron lead to
creation of either three quasiparticles or three quasiholes~\cite{Halp}. In
Fig.1, width of the peaks is approximately inversely proportional to the
magnetic field and, therefore, $\Delta $ is proportional to the peak
amplitude.
\end{thebibliography}
\end{document}